\documentclass[journal]{IEEEtran}
\usepackage{amsmath,amsfonts}
\usepackage{algorithmic}
\usepackage{algorithm}
\usepackage{array}
\usepackage{textcomp}
\usepackage{url}
\usepackage{verbatim}
\usepackage{graphicx}
\usepackage{ntheorem-hyper}
\usepackage{cite}
\usepackage{xcolor}
\newtheorem{lemma}{Lemma}
\newtheorem{remark}{Remark}
\usepackage[top=0.75in,bottom=1in,left=0.625in,right=0.625in]{geometry}
\DeclareMathOperator*{\argmin}{\arg\!\min}
\DeclareMathOperator*{\argmax}{\arg\!\max}

\usepackage{amssymb}

\begin{document}
	\title{Joint Sensing, Communication and Localization of a Silent Abnormality Using Molecular Diffusion}
	
	\author
	{Ladan~Khaloopour, Mahtab~Mirmohseni, \IEEEmembership{Senior Member, IEEE}, \\and Masoumeh~Nasiri-Kenari, \IEEEmembership{Senior Member, IEEE}

		\thanks{This work was supported by the Iran National Science Foundation (INSF) Research Grant on Nano-Network Communications.}
		\thanks{The authors are with the Department of Electrical	Engineering, Sharif University of Technology, Tehran, Iran (e-mail: ladan.khaloopour@ee.sharif.edu; mirmohseni@sharif.edu; mnasiri@sharif.edu).}
	}
	
	\maketitle

	\begin{abstract} 
		In this paper, we propose a molecular communication system to localize an abnormality in a diffusion based medium. We consider a general setup to perform joint sensing, communication and localization. This setup consists of three types of devices, each for a different task: \emph{mobile sensors} for navigation and molecule releasing (for communication), \emph{fusion centers (FC)s} for sampling, amplifying and forwarding the signal, and a \emph{gateway} for making decision or exchanging the information with an external device. The sensors move randomly in the environment to reach the abnormality. We consider both collaborative and non-collaborative sensors that simultaneously release their molecules to the FCs when the number of activated sensors or the moving time reach a certain threshold, respectively. The FCs amplify the received signal and forward it to the gateway for decision making using either an ideal or a noisy communication channel. 
		A practical application of the proposed model is drug delivery in a tissue of human body, in order to guide the nanomachine bound drug to the exact location.
		The decision rules and probabilities of error are obtained for two considered sensors types in both ideal and noisy communication channels.
		
	\end{abstract} 
	


	\section{Introduction}
	
	\IEEEPARstart{M}{olecular} communication (MC) is a new communication paradigm, where the molecules or ions are used as information carriers. MC has been proposed as a promising approach for abnormality detection and localization in
	variety of applications in biotechnology, medicine, and industry; such as air pollutant monitoring, agriculture, abnormality detection and drug delivery \cite{NakanoBook}. 
	For example, the abnormality detection in medical applications has been received increasing attention in recent years \cite{ghavami2017abnormality, mosayebi2017cooperative, nafise2020abnormality,  Varshney2018Vessels, nakano2016performance}. The goal of these works is detecting the virus, bacteria, infectious micro organism or tumors in the body by means of static \cite{ghavami2017abnormality, mosayebi2017cooperative, nafise2020abnormality} or mobile nanomachines \cite{ Varshney2018Vessels, nakano2016performance}. If these abnormalities are detected, the next step would be treatments such as drug delivery, targeted therapy, nanosurgery, regenerative medicine and tissue engineering \cite{felicetti2016applications}. 
	An intermediate step which is necessary between detection and therapy is tracking the target or abnormality localization. This step is crucial to realize the accurate location of the abnormality, in order to eliminate the adverse and side effects of drug on the other normal cells and to enhance the resolution of abnormality before the surgery and other therapies \cite{felicetti2016applications, yangcomprehensive, von2011nanoparticles}. Some examples are localizing the unhealthy cells around the capillaries \cite{Turan2018VesselLike}, guiding the nanorobots in the human body to the target sites needing treatment \cite{Freitas1999nanomedicine, Varshney2018Vessels, Shi2020Nanorobots}, and drug delivery, where the drug can be bounded to the mobile nanomachines to be delivered at the target location \cite{Wang2015Distance, Kumar2020, von2011nanoparticles}.  
	
	In this paper, we focus on the abnormality localization in a two-dimensional (2-D) micro-scale medium, without any physical access to each point of it. We use mobile sensors to transmit the location informations to the outside environment.
	
	\subsection{Related Works}
	
	Using MC setups to localize abnormality has been studied in different scenarios.
	Some research study a scenario, where the target abnormality is a point source who releases molecules into the medium  \cite{Kumar2020, Turan2018VesselLike,  gulec2021fluid}. The goal is to estimate the distance of or localizing this target using receivers, who observe the number of received molecules.  
	In \cite{Kumar2020}, the localization problem is considered in a three-dimensional environment with three receivers. The target is a point source with impulsive release. This work proposes an iterative numerical localization method for two cases of known and unknown receivers locations. \cite{Turan2018VesselLike} considers a vessel-like medium with flow and obtains the abnormality location based on the mean peak concentration of received molecules at the receivers. However, the method of obtaining this mean is not discussed. The authors in \cite{gulec2021fluid} consider a three-dimensional environment where the droplets are used as information carriers. They predict the distance using fluid dynamics algorithm and validate their results with experimental data.
	
	In some scenarios, the target does not release any molecules, but it absorbs the molecules released from a transmitter. Therefore, the concentration of molecules at the receiver is decreased in the presence of target \cite{Kose2020Learning, Bao2021Silent}. 
	The authors in \cite{Kose2020Learning} apply a machine learning method to localize the target using the number of absorbed molecules by the receiver. \cite{Bao2021Silent} proposes an algorithm for target localization based on maximum likelihood estimation (MLE) method.

	In other scenarios, the target either does not release molecules or the concentration of its released molecules becomes very low at the receivers, such that it is detectable only at the vicinity of the target \cite{mosayebi2018early, mosayebi2018advanced}. A solution is to use mobile sensors that reach the target (called navigation) and send their sensing data to receivers by releasing molecules there \cite{nakano2017LFbased, nakano2016leader, okaie2018leader, Khaloopour_sensors_2021}.
	This approach is used both in micro-scale \cite{nakano2017LFbased, nakano2016leader, okaie2018leader} and macro-scale applications \cite{Khaloopour_sensors_2021}.
	The abnormality navigation has two stages.
	In the first stage, the mobile sensors, injected into the environment, are seeking the variations in the medium in order to find the target; this stage is called functional navigation \cite{Freitas1999nanomedicine}. 
	The mobile sensors transfer the sensory data to fusion centers (FC)s, which are used to locate the abnormality. Then, to deliver the drug to the abnormality location, the drug is bound to the nanomachines, who know the abnormality location and walk in the determined route to reach the abnormality \cite{Externally_Controllable, Shi2020Nanorobots , yangcomprehensive}. This stage is called positional navigation \cite{Freitas1999nanomedicine}.
	
	The goal of works on the third scenario is tracking a fixed or mobile target \cite{nakano2017LFbased,  nakano2016leader, okaie2018leader}, or localizing a fixed one \cite{Khaloopour_sensors_2021}. 
	\cite{nakano2017LFbased,  nakano2016leader} consider a 2-D bounded area where two types of nanomachines can move with Brownian motion. When first type reaches the target, it releases molecules to guide a second type nanomachines. The drug is delivered to the target location by second types to decrease the side effects. In \cite{okaie2018leader}, a new type of nanomachines is added to the above model to move randomly and amplify the concentration of molecules in the environment, improving the accuracy of target tracking. 
	In \cite{Khaloopour_sensors_2021}, the detection and localization problems are considered in a macro-scale cylindrical and fluidic environment, with laminar flow condition. The sensors are injected in the medium to detect the silent abnormality and being activated. They also cooperate in activating each other by releasing molecules. The FC localize the abnormality using the sensors status and the received molecules.

	\subsection{Our Contribution}
	We consider a general setup to perform joint sensing, communication, and localization of a silent abnormality in a 2-D environment with no flow, where the abnormality has been previously detected\footnote{We assume that the sensing operation is ideal.}. 
	We consider a target that is identifiable by sensors used in our setup; This is feasible since the target can act 
	{as a source of changes in some environmental parameters, such as temperature, PH level, and concentration of a biomarker in the medium}. In the first case, the sensors interact with the target surface receptors and identify it, and in the second case, the target releases some molecules in the medium, where the sensors can detect them at the vicinity of the releasing point \cite{nakano2017LFbased}. In medical applications, for example drug delivery, this abnormality can be inflammation, tumor, and other unhealthy cells.
	
	\begin{figure}
		\vspace{-1em}
		\centering
		\includegraphics[trim={9.5cm 4cm 9cm 4cm},clip, scale=0.6]{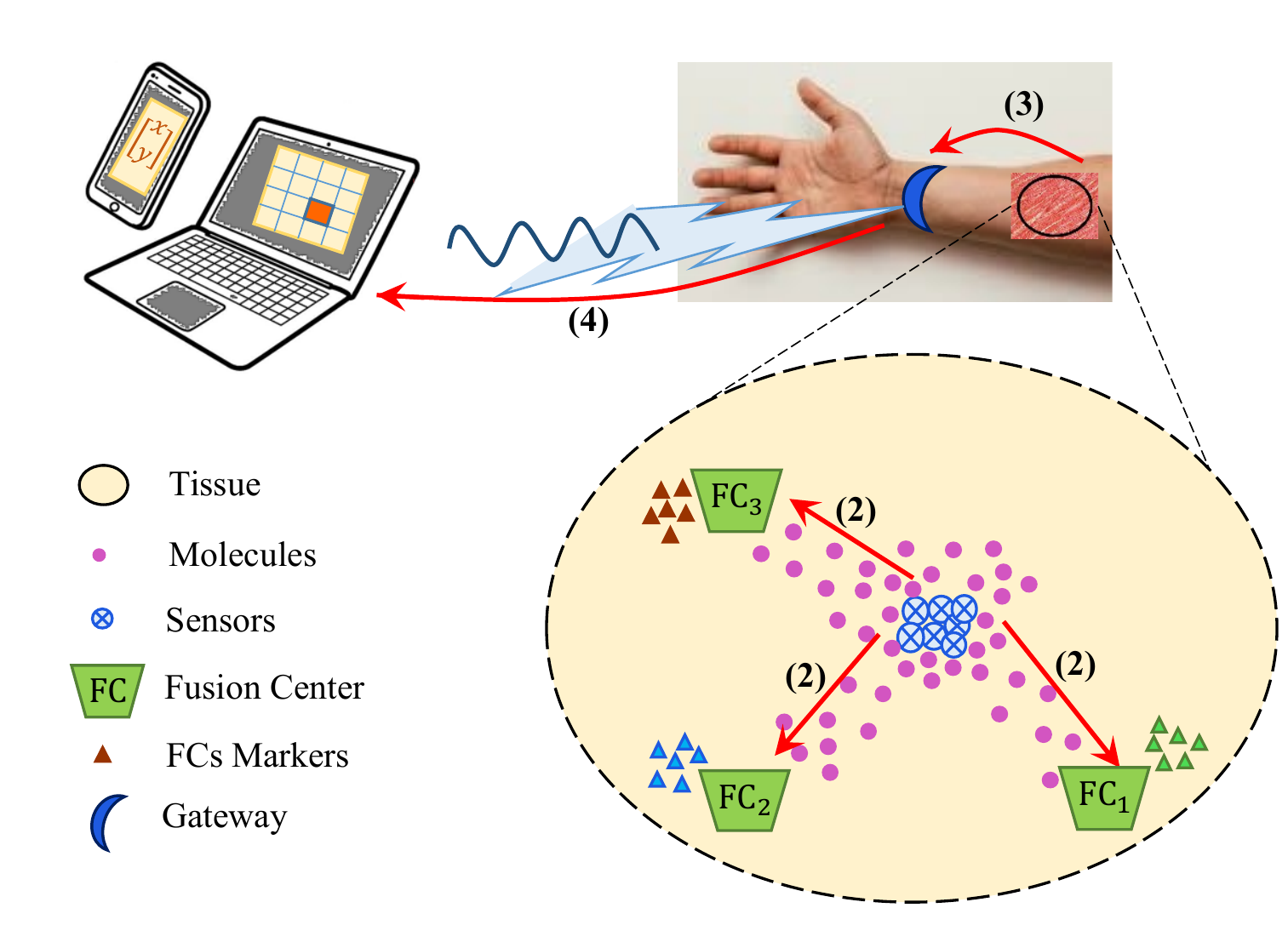}
		\vspace{-1.5em}
		\caption{The Phases (2), (3) and (4) are shown.}
		\label{fig: hand}
		\vspace{0em}
	\end{figure}
	Our localization framework consists of three types of devices used in four phases as shown in Fig. \ref{fig: hand}: some mobile sensors, a few FCs, and a gateway. 
	The sensors can be some biological cells, bacteria, proteins, amino acids, lipids, viruses, or artificial nano-scale machines \cite{nakano2017LFbased, felicetti2016applications}. Phase (1): The sensors {move in the medium randomly}  
	and sense the variations in their vicinity. After sensing an abnormal variation, they stop and get activated. Phase (2): The activated sensors release their stored molecules into the environment after a short delay. {Based on the considered type of sensors (non-collaborative and collaborative), this delay can be deterministic or random.} These molecules are diffused and arrive the FCs located at the vertices of the observing area. Phase (3): Each FC samples the number of received molecules in its volume at some sampling times, amplifies the samples and releases other type of molecules in to the medium.
	These new released molecules will be diffused in the medium to reach the gateway. Phase (4): The gateway may convert the molecular signal into the electromagnetic (EM) wave to be transmitted to an external device \cite{Externally_Controllable, yangcomprehensive} (\emph{i.e.}, computer, laptop, smart phone), or it may have the sufficient computing capabilities to process the molecular signal itself and make decision about the abnormality location. At first, we study the case of ideal communication channel between the FCs and gateway (\emph{i.e.}, noiseless).   
	{For the collaborative sensors, we use two FCs and the abnormality is localized based on its relative distances to the FCs. We obtain the maximum likelihood (ML) decision rule and derive the sub-optimum thresholds to decide the abnormality location. Then, we derive the probability of error in a closed form. For the non-collaborative sensors, we need three FCs for localization. We apply the optimal ML decision rule based on the ratio of FCs observed molecules, which results in a threshold form. Also we derive the probability of error.   
	Then, we study the noisy communication channel between the FCs and gateway. To simplify the equations, we assume that all FCs have equal distances to the gateway. To overcome the difficulty of working with a doubly stochastic random variable (RV), we approximate the random samples of FCs by their mean and obtain the decision rules and probabilities of error.}

	The rest of the paper is organized as follows. In Section \ref{system_model}, the system model is described. Then, the localization problem is investigated for two considered sensors types in Sections \ref{sec: Sensors Strategy 1} and \ref{sec: Sensors Strategy 2}, by obtaining the decision rules and probabilities of error {in the case of ideal channel between the FCs and gateway}. The case of noisy channel is analyzed in Section \ref{sec: Non-ideal Communication Channel}. The numerical and simulation results are provided in Section \ref{sec: numerical}. Finally, the paper is concluded in Section \ref{sec: conclusion}.

	\section{System Model} \label{system_model}

	{We consider a 2-D environment and a $w \times w$ bounded area
	$A=\{[x,y]^ \text T| 0 \leq x \leq {w}, 0 \leq y \leq {w} \}$. We assume that an abnormality exists at an unknown location $\overrightarrow{s_J}=[s_x,s_y]^\text T \in A$ and the location of abnormality is distributed uniformly in $A$. We also assume that the abnormality does not release enough molecules itself\footnote{It may release some molecules but their concentration is very low, detectable only at the vicinity of the abnormality point.}.
	Our goal is to localize the abnormality in some resolution. 
 We divide the area $A$ into some clusters, which are described in the next sections.}
	\begin{figure}
		\vspace{-1em}
		\centering
		\includegraphics[trim={1cm 2cm 10.6cm 4cm},clip, scale=0.4]{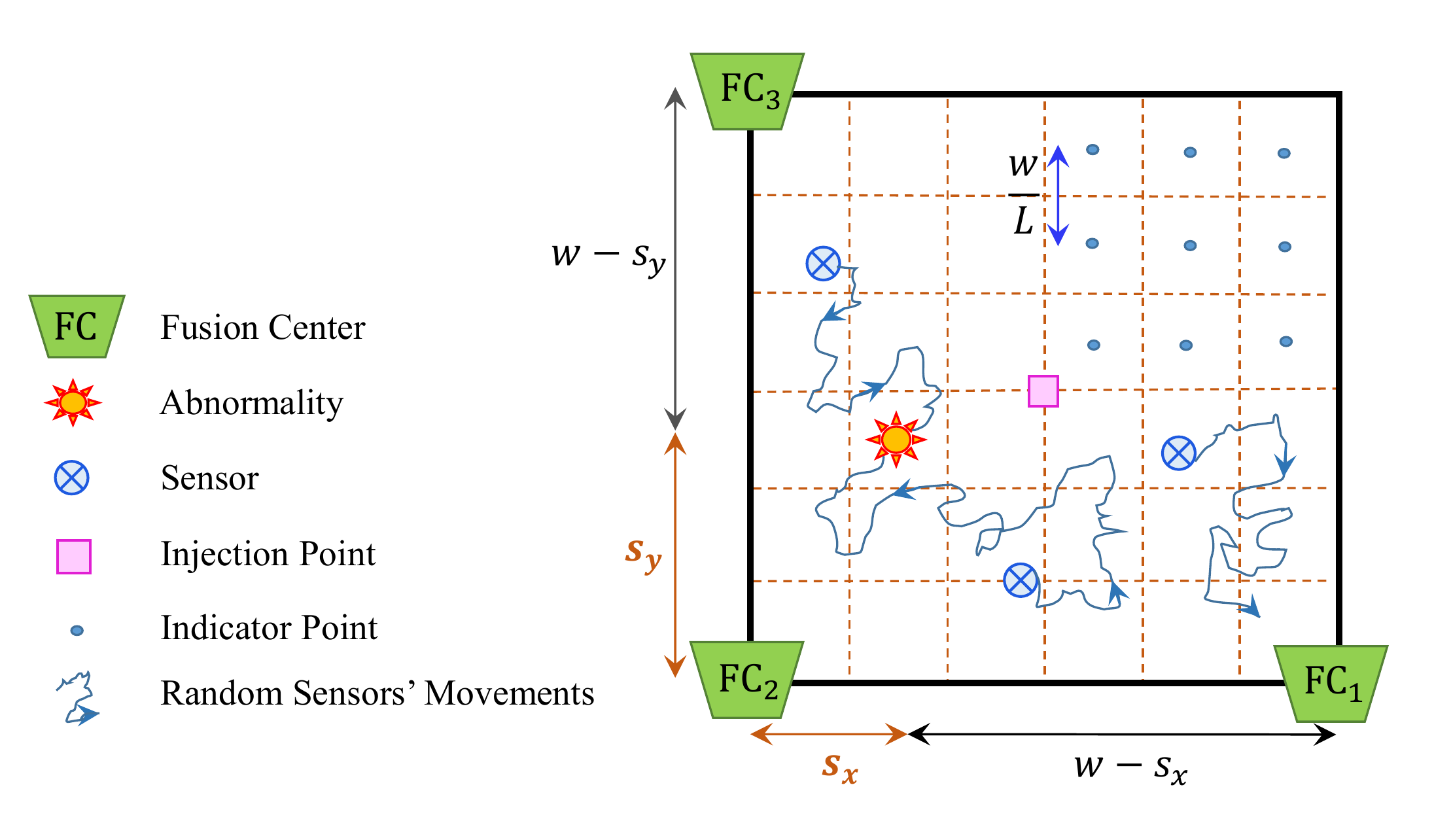}
		\vspace{-1.5em}
		\caption{The bounded observing area, FCs, and sensors random movements to reach the abnormality.}
		\label{fig: Discrete_Points}
		\vspace{0em}
	\end{figure}
	We use three types of devices: some mobile sensors, a few FCs, and a gateway. The functionalities and properties of these devices are described in the following. 
	\\
	$\bullet$ \emph{Mobile sensors}: We utilize $N_s$ mobile sensors, which are injected into the medium at time $t=0$ from the center of the observing area\footnote{or any arbitrary point in the observing area.}, $\overrightarrow{C} = [w/2,w/2]^\text T$. {They can move in the area randomly.}    
	{If a sensor reaches the boundaries, it is reflected back into the area.}  
	The sensors measure an environmental parameter (such as temperature, PH level, and concentration of a biomarker) and are activated by the changes in this parameter at the abnormality location compared with the normal points.
	Each sensor has a molecular storage where it can store $K \geq 1$ types of molecules; $M$ molecules from each type. {Note that all sensors have stored the same $K$ types of molecules.}
	For simplicity, we assume the diffusion coefficient of all these types of molecules in the medium are the same and equal to $D$.
	\\
	$\bullet$ \emph{Fusion centers}: 
	{We use two or three fixed FCs (${\text{FC}_i}$, $i \in \{1,2\}$ or $i \in \{1,2,3\}$) depending on the types of sensors, which will be explained later. The FCs are located at three vertices of the observing area $A$, as shown in Fig. \ref{fig: Discrete_Points}.} 
	Each ${\text{FC}_i}$ has a transparent receiver with the ability of receiving the sensors molecules with volume $V_F$\footnote{For simplicity, we assume that all FCs have equal receiving volumes.}. Each ${\text{FC}_i}$ stores $K_{FC}$ types of molecules, different from sensors molecules. We call the FCs molecules as \emph{markers}. These markers are used to transfer the gathered sensory data to the gateway.
    Each FC includes a transmitter that controls the number of its released markers. An example of this  model is proposed in \cite{Arjmandi2013modulation, Arjmandi2016ionchannel}, where the transmitter has a molecule storage with some surface outlets. The number of released molecules is a Poisson RV whose parameter is determined by the opening size of the outlets \cite{Arjmandi2013modulation}. This size is controlled by a gating parameter signal, which is voltage or ligand concentration in \cite{Arjmandi2016ionchannel}.
\\
$\bullet$ \emph{Gateway}: {The gateway is located far from the observing area; for example for the in-body applications, at a close point to the skin.}  
 It receives the FCs markers in its observing volume $V_G$, at its sampling times. The gateway has larger volume and more computational capabilities compared to the FCs. Thus, it decides the abnormality location itself, or converts its observed samples into an EM wave and transmits it to an external device for decision making. 
 \cite{Hybrid_2019_Deng} proposes an implemented graphene based nano-device, which can convert the molecular signal to an electrical signal. 
 In \cite{Externally_Control_2014}, a smart wristband or other interfaces are used to interconnect the in-body MC and the external environments. For simplicity, we assume that the distance between the gateway and all FCs are equal.

	Using the above three types of devices, our proposed abnormality localization setup consists of four phases. The outline of these phases is as follows. {The sensors move in the observing area.}  When a sensor reaches the abnormality, it gets activated and releases its stored molecules in the medium, which are received by the FCs. They amplify and forward the received molecular signal to the gateway, where either a decision is made or a signal is transmitted to an external device. 	
	{One of the practical applications of this model is drug delivery in a tissue of human body, where the goal is to guide the nanomachine bound drugs to the exact location in order to eliminate the adverse and side effects of drug on the normal cells around the abnormality. Thus, a localization step (\emph{functional navigation}) is necessary before guiding the nanomachines to the target (\emph{positional navigation}).}  
	In the following, these phases are described in more details.
 
	$\bullet$ \textbf{Phase (1)}: 
	{The sensors move randomly in the observing area} and sense the medium, looking for the abnormality by measuring an environmental parameter.
	When a sensor reaches the abnormality location, it gets activated and stops moving. We denote the number of activated sensors by $N_r$.
 
	$\bullet$ \textbf{Phase (2)}:
	 The activated sensors release all of their stored molecules into the medium from the abnormality location in order to be received at the FCs (shown in Fig. \ref{fig: hand}). {The types and number of released molecules are the same for all sensors. These sensors will be degraded in the medium after releasing their molecules.}  
	 We consider two different types of sensors based on their molecule releasing scheme, described as follows. 
	\begin{enumerate}
		\item \textbf{Collaborative sensors}: 
		The sensors used in micro-scale MC are engineered cells, bacteria or artificial nanomachines.  
		{Some engineered bacteria behave socially based on Quorum sensing \cite{bassler1999bacteria, fang2019expected}. It means that they simultaneously respond to an environmental change by cooperating with each other. For example, when the concentration of some signaling molecules around them reaches a critical threshold, they respond together.}   
		{We consider a time-slotted system with slot duration of $T$ and the activated sensors cooperate with each other in signaling time durations of $T_\text{sig}$, where $T_\text{sig} \ll T$. When the number of activated sensors reaches a threshold $N_{th}$, the signaling is finished (we denote this time by $t_r$). The $N_{th}$ sensors release their stored molecules into the medium at the beginning of the next time-slot (at $t=nT$, where $n\in \mathbb{N}$ and $(n-1)T <t_r\leq nT$).}   
		Therefore, in this case, $N_r$ is a fixed number and equals to $N_{th}$. But, the time of this event (releasing time), $t_r$ is not deterministic. 
		For this type of sensors, two FCs (${\text{FC}_i}$, $i \in \{1,2\}$) are used.
		\item \textbf{Non-collaborative sensors}:
		 Other type of sensors, for example artificial nanomachines, have limited energy storage for their mechanical motions (\emph{e.g.}, movements), pumping and etc. This energy may be used for missions with limited durations \cite{Freitas1999nanomedicine}. Thus, the sensors may have limited energy to move for a limited time duration of $T_{th}$. After this time duration, they stop and the activated sensors, which arrive the abnormality, release their stored molecules into the environment. 
		 For the non-collaborative sensors, $N_r$ is an RV.  
		 They release the molecules at the beginning of the next time-slot (\emph{i.e.}, at $t=nT$, where $n\in \mathbb{N}$ and $(n-1)T <T_{th}\leq nT$). So the releasing time of molecules is deterministic. For this type of sensors, three FCs (${\text{FC}_i}$, $i \in \{1,2,3\}$) are used. 
	\end{enumerate}

	$\bullet$ \textbf{Phase (3)}: 
	{The molecules released by the activated sensors diffuse in the medium and some of them reach the FCs. Each FC has $K$ types of receptors to receive all $K$ types of released molecules.} 
	It samples the number molecules in its volume at some sampling times, amplifies the samples and non-instantaneously releases its markers into the medium, where the number of released markers is proportional to its received samples. We remind that each FC has a different type of markers.
	As mentioned before, this number is controlled with a gating parameter, such as voltage or ligand concentration.  
	Unlike the amplification process in \cite{Wang2020RelayAF}, our FCs do not require the full-duplex mode and perfect time synchronization with the sensors. It means that, the release time and duration of the FCs may be different from the sensors. Thus, an FC can send its $K$ samples to the gateway in one of these two ways: (I) Using only one type of markers (\emph{i.e.}, $K_{FC}=1$) and $K$ time-slots to send FC observations to the gateway, or (II) Using $K_{FC}=K$ types of markers in one time-slot. Moreover, the communication channel between the FC and the gateway can be considered ideal (noiseless) or non-ideal (noisy). In the ideal case, there is no need to signal amplification at the FC and the gateway observes the same signals as the FCs.
	
	$\bullet$ \textbf{Phase (4)}: The gateway receives the markers sent by the FCs. Using the number of sampled markers, the abnormality location is decided. We consider that the gateway decides the abnormality location itself instead of an external device. 
	
	{	
	\begin{remark}
		We wish to emphasize the following statements:
		\\
		\textbf{1)} 
		In our system model, we consider a bounded observing area, which is a suitable model for practical applications. We assume that the sensors do not exit this area to guarantee that enough number of sensors reach the abnormality. Without this assumption, we need more sensors. 
		\\
		\textbf{2)} 
		The signaling time duration of collaborative sensors, $T_\text{sig}$, is very short compared with the time intervals between the sensors arriving to the abnormality. Therefore, the probability of that two or more sensors simultaneously arrive the abnormality (and being activated) is negligible. 
		\\
		\textbf{3)} 
		The reason of using $K$ different types of molecules for each sensor is to create $K$ independent observations at the FCs (using $K$ types of receptors). An alternative solution is to use sensors with one molecule type (the total number of molecules is $KM$) and the activated sensors release molecules in $K$ time-slots ($M$ molecules in each time-slot). The FCs receive the molecules in all these $K$ time-slots. Note that the slot duration $T$ is sufficiently large such that the inter symbol interference (ISI) of the previous time-slot is negligible and the observations at each FC are independent.
		\\
		\textbf{4)} 
		For simplicity, we study the localization problem for a 2-D area. This model can also be extended into a 3-D environment easily by utilizing one more FC.  
	\end{remark}}
	
	In the following sections we study the Phases (2) and (3) for Collaborative and Non-collaborative sensors, where the channels between the FCs and the gateway in Phase (3) are ideal (noiseless) or non-ideal (noisy). We derive the decision rules, and obtain the probabilities of error.
	
	\section{Ideal Channel: Collaborative Sensors}  \label{sec: Sensors Strategy 1}
	In this section, we investigate the localization problem for collaborative sensors, with the assumption of ideal channel between FCs and gateway.
	As mentioned, when the number of $N_{th}$ collaborative sensors reach the abnormality location (at time $t_r$), they get activated and each one releases its stored molecules into the environment at $t=nT$, where $n\in \mathbb{N}$ and $(n-1)T <t_r\leq nT$. 
	{Since $t_r$ is not deterministic for the collaborative sensors, $n$ is an RV.
	As each activated sensor releases all its stored molecules ($M$ molecules of each type), the total number of $N_{th} M$ molecules of each type are released by all activated sensors. Note that there are $K$ types of molecules. The FCs sample the number of molecules observed in their volume, at sampling time {$t_s=nT+T_\text{obs}$} where $T_\text{obs}$ is observing time, which is obtained later. We denote the number of $k$-th type of molecules observed at $\text{FC}_i$ by $Y^{k}_i$. 
	Since $n$ in an RV, the FC does not know its realization to decide the sampling time. Thus, it observes the molecules in all time-slots ($n=1,2,\cdots$) until $Y^{k}_i =0$.}  
	If we denote the distance between the ${\text{FC}_i}$ and the abnormality point by $d_i$ (\emph{i.e.}, $d_i \doteq \| \overrightarrow{\text{FC}_i} - \overrightarrow{s_J}\|$), the probability of a molecule to be observed at $t_s$ in the receiver volume at ${\text{FC}_i}$, $i \in {1,2,3}$ is\footnote{We assume that the volume $V_F$ is small enough, such that the concentration of molecules in this volume is uniform.} \cite{Gohari2016survey},
	\begin{align}
	\mu(d_i, T_\text{obs})=\frac{V_F}{{(4\pi D T_\text{obs})^{N/2}}}\exp{(-\frac{d_i^2}{4DT_\text{obs}})}, \label{MU_EQUATION}
	\end{align}
	where $N$ is equal to the dimension of the observing area, which is $N=2$ in our system.
	Each $\text{FC}_i$ obtains $K$ independent samples as $[Y^{1}_i, Y^{2}_i,\cdots, Y^{K}_i]$, by sampling $K$ types of molecules, where $Y^{k}_i \sim \text{Binomial}(N_{th} M,\mu(d_i, T_\text{obs}))$ for $k=1,\cdots,K$. Since the number of released molecules, $M$, is large enough, the Binomial distribution can be approximated by a Gaussian distribution as \cite[p. 105]{papoulis2002},
	\begin{align}
	Y^{k}_i \sim \mathcal{N}(N_{th} M \mu(d_i, T_\text{obs}) , {N_{th} M \mu(d_i, T_\text{obs})}).	\label{Guassian_APX0}
	\end{align}
	The $\mathcal{N}(\eta,\sigma^2)$ in the above equation indicates the Gaussian distribution with mean of $\eta$ and variance of $\sigma^2$.
	The best observing time is the peak time of the molecules concentration, which is obtained from \eqref{MU_EQUATION} as, 
	\begin{align}
	T_\text{obs}=\frac{{d_i}^2}{2ND}.
	\end{align}
	From the above equation, the observing time depends on the distance between the releasing and receiving points of molecules (\emph{i.e.}, $d_i$).
	As mentioned, the releasing point is the abnormality location, which is unknown for the FCs. We choose a sampling time assuming that the abnormality locates at the center of the observing area $A$. Thus, we assume $d_i=\frac{w\sqrt{2}}{2}$ and thus the observing time is $T_\text{obs}=\frac{w^2}{4ND}$. By defining $m(d_i) \overset{\Delta}{=} N_{th} M \mu(d_i, \frac{w^2}{4ND})$, \eqref{Guassian_APX0} results in,
	\begin{align}
	Y^{k}_i \sim \mathcal{N}(m(d_i) , {m(d_i)}).	\label{Guassian_APX}
	\end{align}

	Each $\text{FC}_i$ observes $K$ samples by receiving $K$ molecule types (released from the abnormality point), at the sampling time $t_s$ (\emph{i.e.}, $Y^{k}_i$, for $k=1,\cdots,K$ and $i=1,2$).
	The mean of distribution in \eqref{Guassian_APX}, $m(d_i)$, is a function of $d_i$. Therefore, the distance between $\text{FC}_i$ and the abnormality point ($d_i$) can be obtained by using the number of FCs' sampled molecules.  
	Note that each point in the bounded observing area $A$ can be uniquely presented by its distances to $\text{FC}_1$ and $\text{FC}_2$.
	To localize the abnormality, we cluster the observing area as follows.
	
\begin{figure}
		\vspace{-1em}
		\centering
		\includegraphics[trim={8cm 6cm 7cm 4cm},clip, scale=0.44]{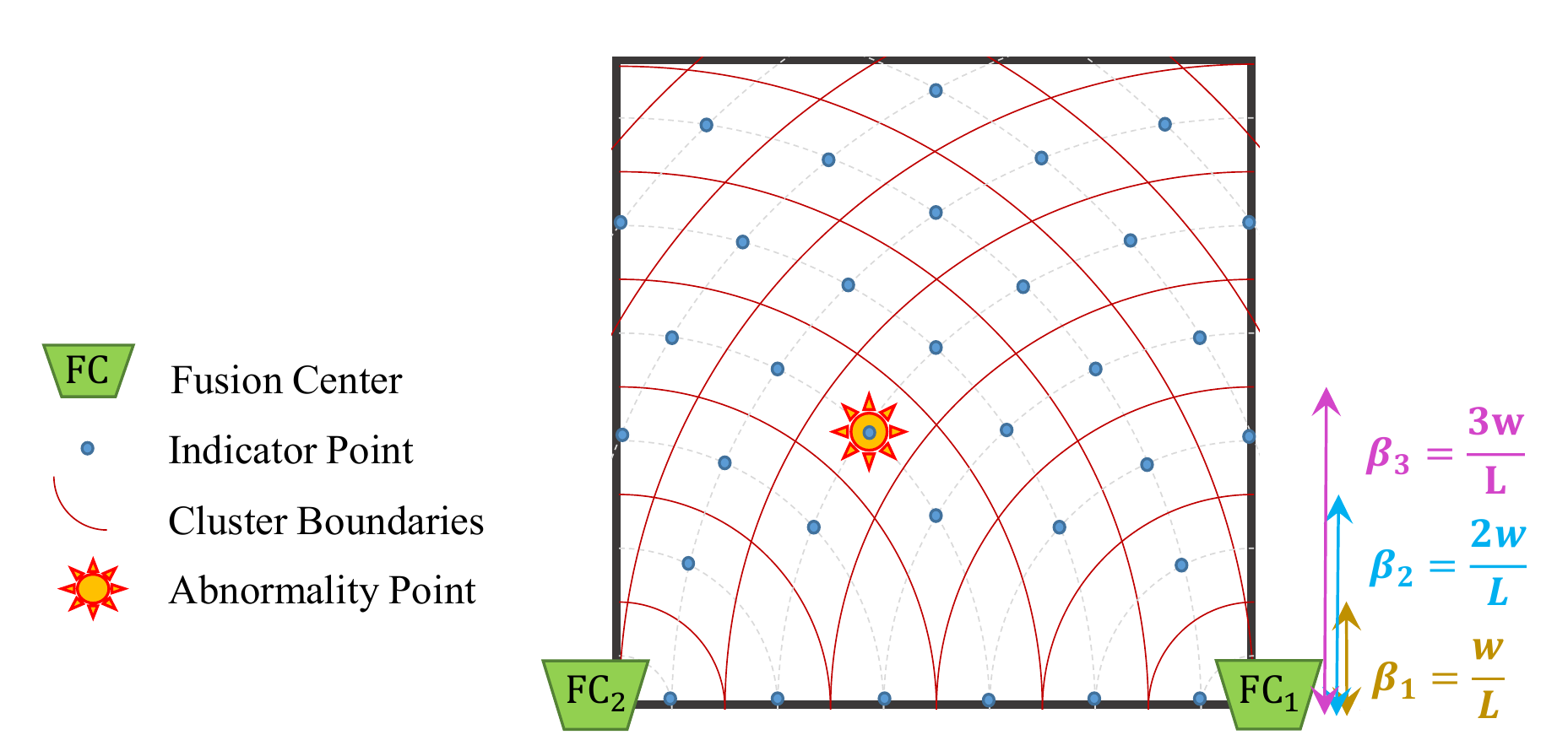}
		\vspace{-0.5em}
		\caption{Clustering the observing area by considering intervals for $d_1$ and $d_2$.}
		\label{fig: radial thresholds}
		\vspace{0em}
	\end{figure}

	$\bullet$ \emph{\textbf{Clustering:}} The abnormality can be located by obtaining the distances $d_1$ and $d_2$. 
	Let $L$ be a clustering parameter and $L \in \mathbb{N}, L\geq 2$. We consider intervals of length $\frac{w}{L}$ for distance $d_i, i\in \{1,2\}$ as  $\{[j\frac{w}{L} \leq d_i < (j+1)\frac{w}{L}]| \forall j=0,1,\cdots\}$, by defining $\beta_j=j\frac{w}{L}$. 
	Therefore, the observing area is partitioned into some clusters as shown in Fig. \ref{fig: radial thresholds} and the localization is performed by deciding the correct cluster, where the abnormality exists. Note that the shape of clusters are different and {we do not know the abnormality distribution in each cluster. Thus, we approximately assume that the abnormality is placed at the center of each cluster, as shown in Fig. \ref{fig: radial thresholds}.} We call these points as indicator points (IP)s, which are placed at distance $d_i \in\{r_j=(j+\frac{1}{2})\frac{w}{L} |j=1,2,\cdots\}$, from $\text{FC}_i$, for $i \in \{1,2\}$.   

\subsection{Decision Rule}
	{Each ${\text{FC}_i}$ amplifies the samples and transmits them to the gateway.} 
	The obtained number of markers by the gateway from ${\text{FC}_i}$ are denoted as $W_i=[W^{1}_i, W^{2}_i,\cdots,W^{K}_i]$, corresponding to $K$ samples of ${\text{FC}_i}$. 
	Note that in this section, we assume that the communication channels between the FCs and gateway are ideal (noiseless). Thus, the gateway observes the same samples as the FCs, \emph{i.e.}, $W^{k}_i = Y^{k}_i$ for $k=1,\cdots, K$. 
	 
	As shown in Fig. \ref{fig: radial thresholds}, the IPs are probable points for the abnormality location. If we show the distances between an IP and $\text{FC}_i$ by $d_i, i \in \{1,2\}$, we denote the set of all IPs distance vectors $(d_1,d_2)$ by $\Psi$, where $|\Psi|=N_p$. 
	We face an $N_p$-hypothesis testing problem, where the decided distance vector specifies the detected cluster.   
	Now, assume that an abnormality exists at an IP with distance vector $(d_1,d_2)$, its location can be found at the gateway by ML decision rule as,
	\begin{align}
	\hat{d_i} &= \argmax_{r_j,j=1,2,\cdots} P(W^{1}_i,\cdots,W^{K}_i|d_i=r_j)   \nonumber \\
	&= \argmax_{r_j,j=1,2,\cdots} P(Y^{1}_i,\cdots,Y^{K}_i|d_i=r_j). 
	\label{ML_rule_1}
	\end{align}
	As mentioned before, $K$ observations are obtained by sampling different types of molecules. Thus, $Y^{1}_i,\cdots,Y^{K}_i$ are independent and thus,
	\begin{align}
	P(Y^{1}_i,\cdots&,Y^{K}_i|d_i=r_j)= \prod_{k=1}^{K}P(Y^{k}_i|d_i=r_j) \nonumber \\
	&\overset{\text{(a)}}{=} \frac{1}{\sqrt{2\pi (m(r_j))^K}}\exp(-\frac{\sum_{k=1}^{K}{(Y^{k}_i-m(r_j))^2}}{2m(r_j)}),
	\nonumber
	\end{align}
	where (a) results from \eqref{Guassian_APX}. 
	By substituting the above equation in \eqref{ML_rule_1} and using some simplification, \eqref{ML_rule_1} can be reduced to the following form.
	\begin{align}
	\hat{d_i} &=\argmin_{r_j,j=1,2,\cdots} K\ln({m(r_j)})+\frac{\sum_{k=1}^{K}{(Y^{k}_i-m(r_j))^2}}{m(r_j)}. \label{ML_RULE_simple}
	\end{align}
	To further simplify the ML decision rule of \eqref{ML_RULE_simple}, we propose a suboptimal and efficient method by comparing two probable IPs with distances $r_{j1}$ and $r_{j2}$ from $\text{FC}_i$ in the following. Using \eqref{ML_RULE_simple}, we have
	\begin{align}
	&K\ln({m(r_{j1})})+\frac{\sum_{k=1}^{K}{(Y^{k}_i-m(r_{j1}))^2}}{m(r_{j1})}  \underset{\hat{d_i} = r_{j1}}{\overset{\hat{d_i} = r_{j2}}{\gtrless}} \nonumber \\ &K\ln({m(r_{j2})})+\frac{\sum_{k=1}^{K}{(Y^{k}_i-m(r_{j2}))^2}}{m(r_{j2})}. 
	\nonumber
	\end{align}
	Noting that $r_{j1}< r_{j2}$, the above decision rule is simplified as,
	\begin{align}
	\sum_{k=1}^{K}(Y^{k}_i)^2 
	\underset{\hat{d_i} = r_{j2}}{\overset{\hat{d_i} = r_{j1}}{\gtrless}}
	Km(r_{j1})m(r_{j2})\Big(1+\frac{\ln(\frac{m(r_{j1})}{m(r_{j2})})}{m(r_{j1})-m(r_{j2})}\Big).
	\label{Decision_Rule_Threshold}
	\end{align}
	The right hand side of \eqref{Decision_Rule_Threshold} is a threshold which depends on distances $r_{j1}$ and $r_{j2}$, defined as 
	\begin{align}
	\tau(r_{j1}, r_{j2})=Km(r_{j1})m(r_{j2})\Big(1+\frac{\ln(\frac{m(r_{j1})}{m(r_{j2})})}{m(r_{j1})-m(r_{j2})}\Big).
	\label{Decision_Threshold_1}
	\end{align}
	Therefore, the proposed suboptimal method compares the sum-squares of samples with a threshold $\tau(r_{j1}, r_{j2})$. In the following section, we derive the probability of error.

\subsection{Probability of Error}

	Assume that the distances of abnormality point from $\text{FC}_1$ and $\text{FC}_2$ are $d_1$ and $d_2$, respectively. If we use the ML decision rule in \eqref{Decision_Threshold_1}, and show the decided distances as $\hat{d_i}$, $i \in \{1,2\}$, the probability of error is
	\begin{align}
	P_e=1-\sum_{(x_1,x_2)\in \Psi}^{}&P[(\hat{d_1},\hat{d_2})=(x_1,x_2)|(d_1,d_2)=(x_1,x_2)]\times
	\nonumber\\
	&P[(d_1,d_2)=(x_1,x_2)]. 
	\label{Pe_strategy_1_ideal}
	\end{align}
	As observations at $\text{FC}_1$ and $\text{FC}_2$ given in \eqref{Guassian_APX} are independent, we have:
	\begin{align}
	p(\hat{d_1},\hat{d_2}|d_1,d_2)=
	p(\hat{d_1}|d_1,d_2)p(\hat{d_2}|d_1,d_2),
	\nonumber
	\end{align}
	where $p(.)$ represent the probability mass function (PMF).
	As the decision on $\hat{d_1}$ only is made based on $d_1$, conditioned on $d_1$, $\hat{d_1}$ is independent of $d_2$. Thus, $p(\hat{d_1}|d_1,d_2)=p(\hat{d_1}|d_1)$. Similarly,  $p(\hat{d_2}|d_1,d_2)=p(\hat{d_2}|d_2)$. Therefore, \eqref{Pe_strategy_1_ideal} results in
	\begin{align}
	P_e=1-\frac{1}{N_p}\sum_{(x_1,x_2)\in \Psi}^{} \prod_{i=1,2} P[\hat{d_i}=x_i|d_i=x_i],
	\label{P_e_total_scheme_1}
	\end{align}
	where from \eqref{Decision_Threshold_1} we have,
	\begin{align}
	&P[\hat{d_i}=x_i|d_i=x_i]= \label{Pe_Sum_Square_1}\\
	&P[\tau(x_i+\frac{w}{L},x_i)<\sum_{k=1}^{K}(Y^{k}_i|d_i=x_i)^2 < \tau(x_i,x_i-\frac{w}{L})].
	\nonumber
	\end{align}
	To simplify \eqref{Pe_Sum_Square_1}, we obtain the distribution of $\sum_{k=1}^{K}(Y^{k}_i|d_i=x_i)^2$. The $(Y^{k}_i|d_i=x_i)$ has a Gaussian distribution as $\mathcal{N}(m(x_i) , {m(x_i)})$, where its mean is not zero.  
	Now we write \eqref{Pe_Sum_Square_1} as,
	\begin{align}
	&P[\hat{d_i}=x_i|d_i=x_i]= \label{Pe_Sum_Square_3}\\
	&P[\frac{\tau(x_i+\frac{w}{L},x_i)}{m(x_i)}<\sum_{k=1}^{K}\big(\frac{Y^{k}_i|d_i=x_i}{\sqrt{m(x_i)}}\big)^2 < \frac{\tau(x_i,x_i-\frac{w}{L})}{m(x_i)}].
	\nonumber
	\end{align}
	All $Y^{k}_i|d_i=x_i$s have the same variance of ${m(x_i)}$. 
	Therefore, $\sum_{k=1}^{K}\big(\frac{Y^{k}_i|d_i=x_i}{\sqrt{m(x_i)}}\big)^2$ has a non-central Chi-squared distribution, using Lemma \ref{lemma_1} in Appendix \ref{lemmas_1_and_2}. Thus, we write \eqref{Pe_Sum_Square_3} as
	\begin{align}
	P[\hat{d_i}=x_i&|d_i=x_i]=Q_{\frac{K}{2}}(\sqrt{K m(x_i)} ,\sqrt{\frac{\tau(x_i,x_i+\frac{w}{L})}{m(x_i)}})
	 \nonumber \\
	&-Q_{\frac{K}{2}}(\sqrt{K m(x_i)} ,\sqrt{\frac{\tau(x_i,x_i-\frac{w}{L})}{m(x_i)}}).
	\label{P_d_Chi_1}
	\end{align}
	The probability of localization error is obtained by substituting \eqref{P_d_Chi_1} in \eqref{P_e_total_scheme_1}.

\section{Ideal Channel: Non-collaborative Sensors}  \label{sec: Sensors Strategy 2}
	
	In this section, we investigate the localization problem for non-collaborative sensors, with the assumption of ideal channels between FCs and gateway. Thus, the gateway observes the same samples as the FCs (\emph{i.e.}, $W^{k}_i = Y^{k}_i$ for $k=1,\cdots, K$ and $i \in \{1,2,3\}$). 
	 We denote the number of non-collaborative sensors that reach the abnormality location at the stop time by $N_r$, which is an RV. These sensors release their stored molecules into the environment at $t=nT$, where $n \in \mathbb{N}$ and $(n-1)T <T_{th}\leq nT$. So $N_r M$ molecules of each type are released, and the FCs'  sampling times are deterministic as $t_s=nT+T_\text{obs}$. The number of received molecules is a Gaussian RV as \eqref{Guassian_APX}, where $m(d_i)=M N_r\mu(d_i, T_\text{obs})$ and $i \in \{1,2,3\}$, in this section. Our goal is to find the abnormality location $\overrightarrow{s_J}=[s_x,s_y]^\text T$. As illustrated in Fig. \ref{fig: Discrete_Points}, we have
	\begin{align}
	d_1^2&=\| \overrightarrow{\text{FC}_1} - \overrightarrow{s_J}\|^2=(w-s_x)^2+s_y^2,
	\nonumber
	\\
	d_2^2&=\| \overrightarrow{\text{FC}_2} - \overrightarrow{s_J}\|^2=s_x^2+s_y^2,
	\nonumber
	\\
	d_3^2&=\| \overrightarrow{\text{FC}_3} - \overrightarrow{s_J}\|^2=s_x^2+(w-s_y)^2.
	\nonumber 
	\end{align}
	
	$\bullet$ \emph{\textbf{Clustering:}}
	In this section, we localize the abnormality by deciding the $s_x$ and $s_y$. These variables are dependent. We consider intervals of length $\frac{w}{L}$ for $s_x$ and $s_y$, as $\{[j\frac{w}{L} \leq s_x , s_y < (j+1)\frac{w}{L}]| \forall j=0,1,\cdots, L-1\}$.  
	Therefore, the observing area is partitioned into some clusters as shown in Fig. \ref{fig: Discrete_Points}, and the localization is performed by deciding the correct cluster, where the abnormality exists. Note that in this case, the shape of clusters are the same. Therefore, the abnormality location is distributed uniformly in each cluster. We approximately assume that the abnormality would be located at the IPs placed at $[(i-\frac{1}{2})\frac{w}{L},(j-\frac{1}{2})\frac{w}{L}]^\text{T}$, where $i,j\in\{1,\cdots,L\}$ as shown in Fig. \ref{fig: Discrete_Points}.
	
	Our approach is to find $s_x$ using the observations of $\text{FC}_1$ and $\text{FC}_2$, and $s_y$ using the observations of $\text{FC}_2$ and $\text{FC}_3$. First, we focus on finding $s_x$. The procedure for $s_y$ is similar. We 
	convert $K$ samples into one sample by averaging, and define a new RV as,
	\begin{align}
	V_i &\doteq \frac{1}{K}\sum_{k=1}^{K} W^k_{i} = \frac{1}{K}\sum_{k=1}^{K} Y^k_{i} \nonumber \\
	&\sim \mathcal N (m(d_i), \frac{1}{K} m(d_i)), \text{ } i \in\{1,2\}.\nonumber  
	\end{align}
	If $\sqrt{Km(d_i)} \geq 10$, we use Lemma \ref{lemma_2} and define a positive RV as $Z_{12} = \frac{V_1}{V_2}$, which has approximately a normal distribution as $Z_{12}\sim \mathcal N (\mu_{Z_{12}},\sigma^2_{Z_{12}})$, where
		\begin{align}
		\mu_{Z_{12}}&{=}\frac{m(d_1)}{m(d_2)}=\exp(\frac{2w s_x-w^2}{4DT_\text{obs}}), \label{mean_Z12} 
		\\
		\sigma^2_{Z_{12}}&{=}\frac{m(d_1)}{Km^2(d_2)}\big(1+\frac{m(d_1)}{m(d_2)}\big)= \frac{\mu_{Z_{12}}}{Km(d_2)}\big(1+\mu_{Z_{12}}\big).
		\label{sigma2_Z12} 
		\end{align}
	Note that the mean and variance of $Y^k_{i}$s, $i \in \{1,2,3\}$ depend on $N_r$, which is an RV. But the mean of $\mu_{Z_{12}}$ is deterministic and does not depend on $N_r$.  
The variance $\sigma^2_{Z_{12}}$ in \eqref{sigma2_Z12} depends on $m(d_2)$, which is the mean number of received molecules by $\text{FC}_2$. It is an RV and its realization is unknown at $\text{FC}_2$. In the following, we approximate $m(d_2)$ to find $\sigma^2_{Z_{12}}$.

{We estimate $m(d_2)$ using the observations $Y^k_2$, $k\in\{1,\cdots,K\}$; $m(d_2)$ is chosen as the minimizer of the mean square error (MSE)\footnote{If $K$ is large enough and $\mathbf{E}[.]$ be the expectation operator, the MSE approximation tends to $\mathbf{E}[Y_2]=m(d_2)$ (law of large numbers (LLN)).}, 
	\begin{align}
	m(d_2) \simeq\argmin_{m} \sum_{k=1}^{K}(m-Y_2^k)^2 = \frac{1}{K}\sum_{k=1}^{K}Y_2^k.
	\nonumber
	\end{align}	
We use the above approximation for designing the system and deriving the decision rules, 
and it does not effect the performance analysis of the system (\emph{i.e.}, the probability of error derivations).}

\subsection{Decision Rule}
	
	To decide the abnormality location ($s_x$ and $s_y$), we define $i_x=s_x\frac{L}{w}+\frac{1}{2}$ 
	 and 
	  $i_y=s_y\frac{L}{w}+\frac{1}{2}$,
	where $i_x$ and $i_y$ are independently and uniformly distributed over $ \{1,\cdots,L\}$ (\emph{i.e.}, $p(i_x)=p(i_y)=\frac{1}{L}$). Thus, the optimal ML decision rule is 
	\begin{align}
	\hat{i}_x &= \argmax_{r_j,j=1,2,\cdots} P[Z_{12}=z_{12}|i_x],
	\nonumber
	\end{align} 
	where $z_{12}$ denotes the realization of $Z_{12}$.
	To find a threshold based decision rule, we consider the log-likelihood ratio (LLR) as,
	\begin{align}
	\Lambda(z_{12}) =  \ln(\frac{P[{Z_{12}}={z_{12}}|i_x]}{P[{Z_{12}}={z_{12}}|i_x+1]}). 
	\label{llr_ML}
	\end{align}
	Using the normal distribution of ${Z_{12}}$, \eqref{llr_ML} will be reduced into
	\begin{align}
	\Lambda({z_{12}}) =&
	\ln\big(\frac{\sigma_{Z_{12}|i_x+1}}{\sigma_{Z_{12}|i_x}}\big) 
	+\frac{(z_{12}-\mu_{Z_{12}|i_x+1})^2}{2\sigma^2_{Z_{12}|i_x+1}}
\nonumber\\
&-\frac{(z_{12}-\mu_{Z_{12}|i_x})^2}{2\sigma^2_{Z_{12}|i_x}},
	\nonumber
	\end{align}
	which can be simplified as,
	\begin{align}
	\Lambda({z_{12}}) =
	a z^2_{12}+ b z_{12}+c,
	\nonumber
	\end{align}
	where
	\begin{align}
	& a = \frac{1}{2}(\frac{1}{\sigma^2_{Z_{12}|i_x+1}}-\frac{1}{\sigma^2_{Z_{12}|i_x}}),
	\nonumber \\
	& b = \frac{\mu_{Z_{12}|i_x}}{\sigma^2_{Z_{12}|i_x}}-\frac{\mu_{Z_{12}|i_x+1}}{\sigma^2_{Z_{12}|i_x+1}},
	\nonumber \\
	& c = \frac{1}{2} \left(\ln(\frac{\sigma^2_{Z_{12}|i_x+1}}{\sigma^2_{Z_{12}|i_x}})
	+\frac{\mu^2_{Z_{12}|i_x+1}}{\sigma^2_{Z_{12}|i_x+1}}
	-\frac{\mu^2_{Z_{12}|i_x}}{\sigma^2_{Z_{12}|i_x}} \right).
	\nonumber
	\end{align}
	
	We denote the optimal threshold between $i_x$ and $i_x+1$ by $\gamma(i_x)$, which is obtained by solving $\Lambda({\gamma(i_x)})=0$, and the optimum decision rule is $\Lambda({z_{12}}) \underset{i_x+1}{\overset{i_x}{\gtrless}} 0$. Noting that $a<0$, the decision rule is obtained as
	\begin{align}
	z_{12}
	\underset{i_x}{\overset{i_x+1}{\gtrless}}
	\gamma(i_x)=-\frac{b}{2a} \overset{+}{-}\sqrt{(\frac{b}{2a})^2-\frac{c}{a}}.
	\label{ml_rule_simplified_2}
	\end{align}

	So far, we have discussed the decision rule to find $\hat s_x=\hat i_x\frac{w}{L}$. Note that the $\hat s_y=\hat i_y\frac{w}{L}$ can also be decided similarly by considering the $Z_{32}=\frac{X_3}{X_2}$. Then the location of abnormality in the observing area is obtained as $\overrightarrow{s_J}=[\hat s_x , \hat s_y]^\text T$.

\subsection{Probability of Error}
An error occurs when $\{\hat s_x \neq s_x\}$ or $\{\hat s_y \neq s_y\}$. Thus, we have
\begin{align}
P_e &= P[\hat s_x \neq s_x \cup \hat s_y \neq s_y] \nonumber \\
& = \sum_{i_x=1}^{L}
\sum_{i_y=1}^{L} P[\hat s_x \neq s_x \cup \hat s_y \neq s_y|(i_x, i_y)]P[(i_x, i_y)], \label{Pe_strategy_2}
\end{align}
where $P[(i_x, i_y)]=\frac{1}{L^2}$.
Using \eqref{ml_rule_simplified_2}, we have
\begin{align}
&P[\hat s_x \neq s_x \cup \hat s_y \neq s_y|(i_x, i_y)]
 \nonumber\\
&= 1-P[\hat s_x = s_x \cap \hat s_y = s_y| i_y] \nonumber\\
&= 1-P[\gamma(i_x-1)<z_{12}<\gamma(i_x)]P[\gamma(i_y-1)<z_{32}<\gamma(i_y)]. \nonumber
\end{align}
Since $Z_{12}\sim \mathcal N (\mu_{Z_{12}},\sigma^2_{Z_{12}})$, we have
\begin{align}
P[\gamma(i_x-1)<&z_{12}<\gamma(i_x)] \nonumber  \\ &=Q(\frac{\gamma(i_x-1)-\mu_{Z_{12}}}{\sigma_{Z_{12}}}) -Q(\frac{\gamma(i_x)-\mu_{Z_{12}}}{\sigma_{Z_{12}}}). \nonumber
\end{align}
$P[\gamma(i_y-1)<z_{32}<\gamma(i_y)]$ can be also obtained similarly.

\section{Non-ideal Communication Channel} 
\label{sec: Non-ideal Communication Channel}
In this section, we consider the noise of communication channel between the FCs and the gateway, which is located at $\overrightarrow{s_G}$. 
Each FC samples the number of received molecules in its volume at the sampling times, amplifies the samples and non-instantaneously release other type of molecules (called as markers) into the medium, where the number of released markers is proportional to the received samples. The FC has a transmitter that controls the number of released markers. An example of this model is proposed in \cite{Arjmandi2013modulation, Arjmandi2016ionchannel} and used in \cite{Mosayebi2018typebased}, where the transmitter has a molecule storage with surface outlets. The number of released molecules is a Poisson RV. The parameter of this variable is determined by the opening size of the outlets \cite{Arjmandi2013modulation}.  In \cite{Arjmandi2016ionchannel}, this size is controlled by gating parameter signal (\emph{i.e.,} voltage, ligand concentration).

 We assume that each FC uses different types of markers to amplify and forward its received signal to the gateway. The gateway is a transparent receiver and has the receptors of all types of markers. Thus, it receives all types of markers, independently. For simplicity, we assume that the distances between the FCs and the gateway are the same and equal to $d_{FG}$ and the diffusion coefficients for all types of markers used by FCs are the same and equal to $D_2$. 

In the following subsections, we investigate the performance of collaborative and non-collaborative sensors, in the presence of noisy channel between the FCs and the gateway.

\subsection{Collaborative Sensors}	 \label{sec: RG_1}
 
Each FC obtains $K$ independent samples, amplifies and forwards the observed signals with $K$ different markers. Focusing on the $k$-th type of markers, the number of released markers from ${\text{FC}_i}$ is $\alpha Y^k_i$, where $\alpha \in \mathbb{N}$ is the amplification factor.   
The probability of a marker released by ${\text{FC}_i}$ be observed after duration of $T_{G,\text{obs}}$ in gateway's receptor volume is\footnote{We assume that the volume $V_G$ is small enough, such that the concentration of markers in this volume is uniform.} \cite{Gohari2016survey},
\begin{align}
\tilde\mu(d_{FG}, T_{G,\text{obs}})=\frac{V_{G}}{{4\pi D_2 T_{G,\text{obs}}}}\exp{(-\frac{d_{FG}^2}{4D_2T_{G,\text{obs}}})}. \label{MU_EQUATION_RG}
\end{align}
The number of markers observed by the gateway is an RV as $W^k_i \sim \text{Binomial}(\alpha Y^k_i,\tilde\mu(d_{FG}, T_{G,\text{obs}}))$. Since the number of released markers, $\alpha Y_i^k$, is large enough, the Binomial distribution can be approximated by Gaussian {as \cite[p. 105]{papoulis2002}}, 
\begin{align}
W^{k}_i \sim \mathcal{N}(\alpha Y_i^k \tilde\mu(d_{FG}, T_{G,\text{obs}}) , {\alpha Y_i^k  \tilde\mu(d_{FG}, T_{G,\text{obs}})}).	\label{Guassian_APX0_RG}
\end{align}
The best observing time for the gateway is the peak time of the markers concentration in \eqref{MU_EQUATION_RG}, which is obtained as
$
T_{G,\text{obs}}=\frac{{d^2_{RG}}}{4D_2}
$.
Defining $m_G(d_i) \overset{\Delta}{=}\alpha Y_i^k \tilde\mu(d_{FG}, \frac{{d_{FG}}^2}{2ND_2})$, \eqref{Guassian_APX0_RG} results in
\begin{align}
W^{k}_i \sim \mathcal{N}(m_G(d_i) , {m_G(d_i)}).	\label{Guassian_APX_FG}
\end{align}
Due to \eqref{Guassian_APX} and \eqref{Guassian_APX0_RG}, RVs $d_i$, $Y^k_i$, and $W_i^k$ form a \emph{Markov chain} $d_i~\rightarrow~Y^k_i~\rightarrow~W_i^k$. Note that the mean number of observed markers at the gateway (\emph{i.e.}, $m_G(d_i)$) is an RV and the $W^{k}_i$ is a doubly stochastic RV. 
The ML decision rule at gateway is 
\begin{align}
\hat{d_i} &= \argmax_{d} P(W_i|d) 
= \argmax_{d} \prod_{k=1}^{K}P(W^k_i|d), \nonumber 
\end{align}
where
\begin{align}
&P(W^k_i|d) = \int_{y=0}^{\infty} P(W^k_i|Y_i^k=y,d)P(Y_i^k=y|d)dy,
\label{MAP_Rule_Non_Ideal} 
\\
 &P(W^k_i=w|Y_i^k=y,d)= P(W^k_i=w|Y_i^k=y)= 
 \nonumber
  \\
 &\big(2\pi (m_G(d_i)|Y_i^k=y)\big)^{-\frac{1}{2}}\exp\big(-\frac{(w- (m_G(d_i)|Y_i^k=y)))^2}{2(m_G(d_i)|Y_i^k=y)}\big), \nonumber 
  \\
 & P(Y_i^k=y|d)=\label{normal_dist_Y|d} \\
 &\big(2\pi  N_{th} M \mu(d, \frac{w^2}{4ND})\big)^{-\frac{1}{2}}\exp(-\frac{(y- N_{th} M \mu(d, \frac{w^2}{4ND}))^2}{2  N_{th} M \mu(d, \frac{w^2}{4ND})}). \nonumber
\end{align}
{$W^{k}_i$ is a doubly stochastic RV and as can be seen, \eqref{MAP_Rule_Non_Ideal} cannot be derived in a closed form.
To make the analyzes tractable, one approach is to approximate the mean of $W^{k}_i$ with a deterministic variable \cite{Ahmadzadeh_2015_AF_Relay, cheng2021amplify}. We call this method as mean-value approximation. 
Thus, we approximate $m_G(d_i)$ with its mean, $\text{E}(m_G(d_{i}))=\alpha \text{E}(Y_i^k)\tilde\mu(d_{FG}, T_{G,\text{obs}})$. Therefore, $Y_i^k \approx m(d_i)$, and $m_G(d_i) \approx \alpha m(d_i)  \tilde\mu(d_{FG}, T_{G,\text{obs}})$, (this is equivalent to approximating the probability distribution function (PDF) of normal distribution in \eqref{normal_dist_Y|d} with a Dirac delta function). We obtain the sub-optimal decision rule similar to Section \ref{sec: Sensors Strategy 1}, where the thresholds will be obtained from \eqref{Decision_Threshold_1} by substituting $m(r_{j1})$ and $m(r_{j1})$ with $\text{E}(m_G(r_{j1}))$ and $\text{E}(m_G(r_{j1}))$, respectively. 
The accuracy of the above approximation is confirmed in Section \ref{sec: numerical}.} 
  
{The probability of error is obtained as \eqref{P_e_total_scheme_1}
where 
\begin{align}
P[\hat{d_i}=x_i|d_i=x_i]&= \sum_{y=0}^{\infty}P[ Y_i^k=y|d_i=x_i] \times 
\nonumber
\\
P\Big[\tau(x_i+\frac{w}{L},x_i)<&\sum_{k=1}^{K}(W^{k}_i|d_i=x_i, Y_i^k=y)^2 < \tau(x_i,x_i-\frac{w}{L})\Big] 
\nonumber
\\
\simeq 
\Big[\tau(x_i+\frac{w}{L},x_i)<&\sum_{k=1}^{K}(W^{k}_i| Y_i^k=m(d_i))^2 < \tau(x_i,x_i-\frac{w}{L})\Big].
\nonumber
\end{align}
}

\subsection{Non-collaborative Sensors}	
As stated in Section \ref{sec: Sensors Strategy 2}, we approximate the ratio of FCs observations as a normal distribution,
in order to analyze the system performance. We define $Z_{12} = \frac{V_1}{V_2}$, where $V_i = \frac{1}{K}\sum_{k=1}^{K} W^k_{i} \sim \mathcal N (m_G(d_i), \frac{1}{K} m_G(d_i))$, for $i\in\{1,2,3\}$. Using Lemma \ref{lemma_2}, we have
\begin{align}
Z_{12} \sim \mathcal N \Big(\frac{m_G(d_1)}{m_G(d_2)},
\frac{m_G(d_1)}{m_G^2(d_2)}\big(1+\frac{m_G(d_1)}{m_G(d_2)}\big) \Big).
\nonumber
\end{align}
The $W^{k}_i$s and thus $V_i$s are doubly stochastic RVs. Similar to Subsection \ref{sec: RG_1}, we approximate the observed $Y^k_i$ by its mean as $Y^k_i=m(d_i)$. Thus, the decision rule and the probability of error are obtained as \eqref{ml_rule_simplified_2} and \eqref{Pe_strategy_2} by substituting $m(d_i)$ with $m_G(d_i)\simeq \alpha m(d_i)\tilde\mu(d_{FG}, T_{G,\text{obs}})$. 

\section{Numerical and Simulation Results} \label{sec: numerical}
\textbf{System Parameters:}
	The system parameters are shown in Table \ref{table: sys_param}. In order to have a fair comparison of both types of sensors, we assume that at stop time, $T_{th}$, the number of activated  
	non-collaborative sensors is $N_{th}$ (\emph{i.e.}, $N_r=N_{th}$). Thus, for both types of collaborative and non-collaborative sensors, the number of activated sensors, which release molecules, are equal. 
	Note that in the case of non-collaborative sensors, the FCs do not know the number of activated sensors. 
	\begin{table}
		\centering
		\caption{System Parameters}
			\begin{tabular}{ c l l}
				\hline Symbol & Description  & Value \\\hline 
				$N$ & number of environment dimensions & $2$ \\
				$D$ & Diffusion coefficient of molecules & $10^{-9}$ \\
				$w$ & Width of area & $10^{-2}$ \\
				$K$ & Number of samples & $2$ \\
				$N_{th}M$ & Number of released molecules of each type & $10^{6}$ \\
				$V_F$ & Observing volume of $\text{FC}_i$ & $1.11\times10^{-7}$ \\
				$D_2$ & Diffusion coefficient of the FCs markers & $10^{-10}$ \\
				$V_{G}$ & Observing volume of the gateway & $1.78\times10^{-6}$ \\
				$\alpha$ & Amplification factor at the FCs & $10^3$ \\
				\hline
			\end{tabular}
			\label{table: sys_param}
			\vspace{-0.5em}
	\end{table}
\begin{figure}
	\vspace{-1em}
	\centering
	\includegraphics[trim={1cm 0cm 1cm 0cm},clip, scale=0.47]{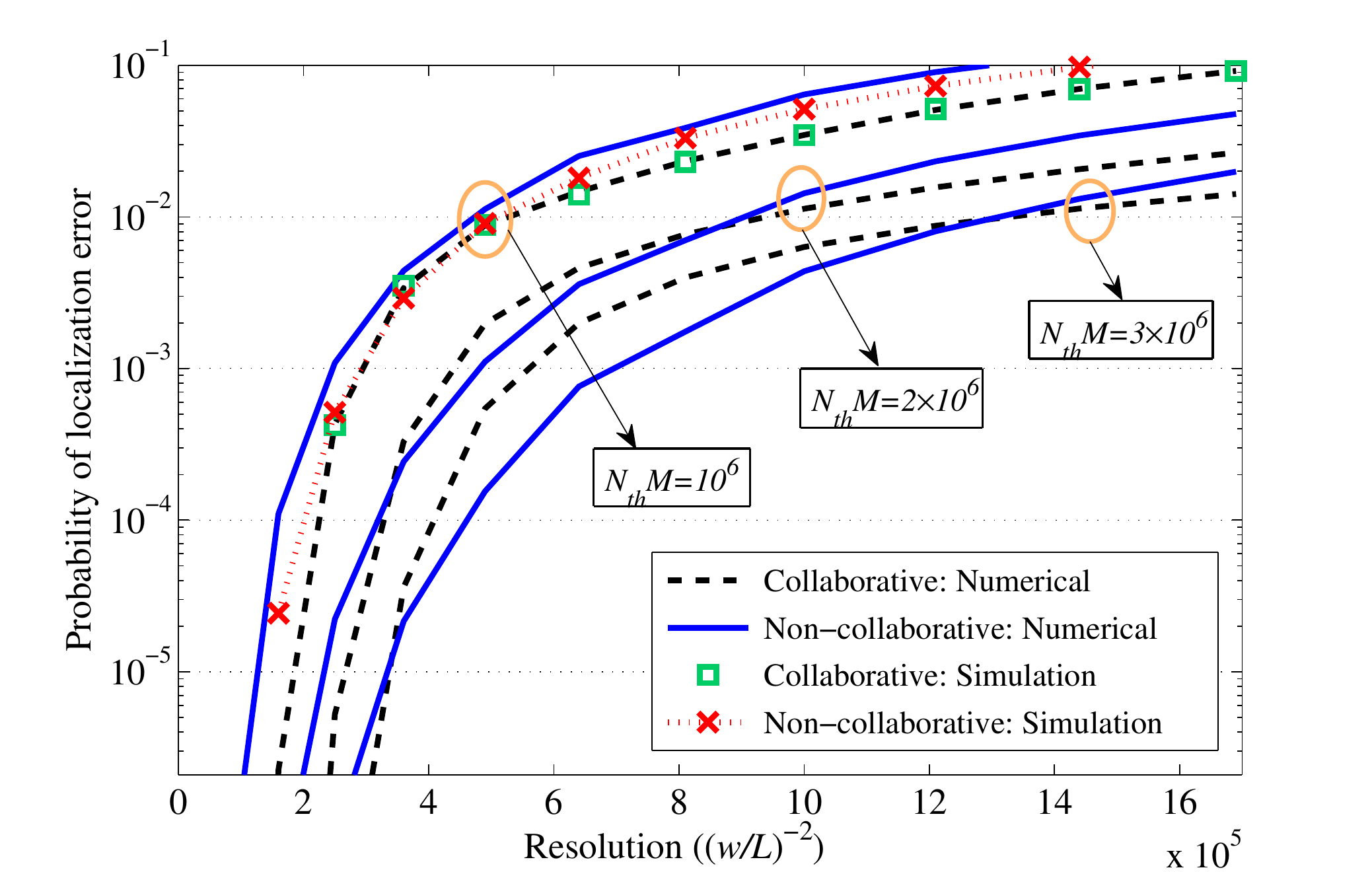}
	\caption{The probability of error versus the resolution of localization, for ideal communication channel between FCs and gateway.}
	\label{fig: result 1}
	\vspace{0em}
\end{figure}


	First, we define the localization resolution as the inverse of sub-region area in Fig. \ref{fig: Discrete_Points} as $(\frac{w}{L})^{-2}$. 
	The probability of localization error versus the resolution is shown in Fig. \ref{fig: result 1}.
	Note that for the cases of collaborative and non-collaborative sensors, two and three FCs are used, respectively. As can be seen, if the number of released molecules is $N_{th}M=10^6$, the performance of collaborative sensors is better than non-collaborative ones. Also we can see that for higher number of molecules ($N_{th}M=2\times 10^6, 3\times 10^6$), the non-collaborative sensors performs better than the collaborative ones for the medium resolutions, as it uses one more FC for deciding the location. It is also observed that the probability of localization error decreases if the total number of molecules increases. The simulation results are also provided in Fig. \ref{fig: result 1} for the case of $M=10^6$, which perfectly validate the numerical results for collaborative sensors. The gap between the numerical and simulation results for non-collaborative sensors is due to the approximation used in Section \ref{sec: Sensors Strategy 2} for the ratio of two normal distributions.

\begin{figure}
	\vspace{-1em}
	\centering
	\includegraphics[trim={1cm 0cm 1cm 0cm},clip, scale=0.47]{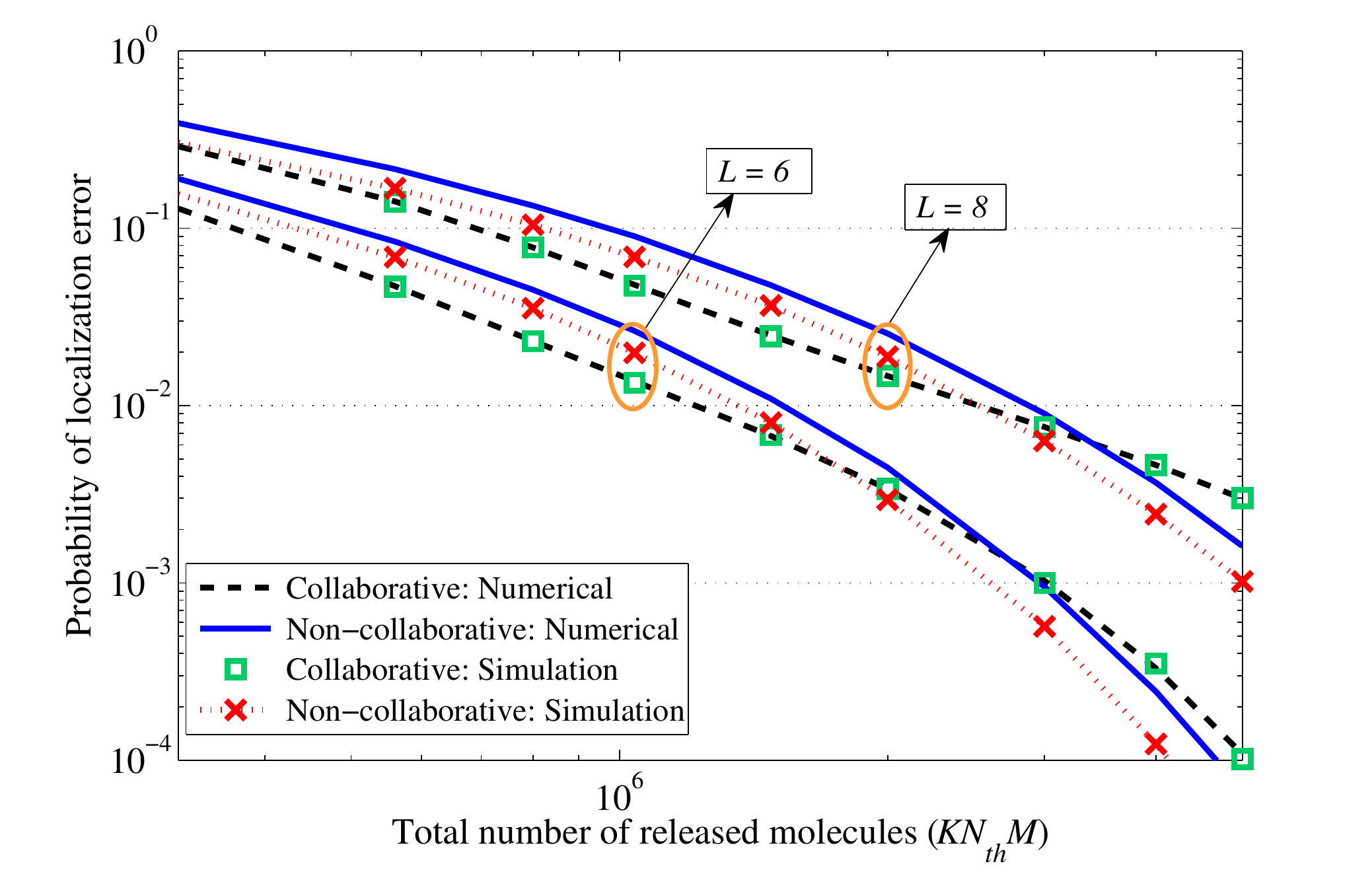}
	\vspace{-1.5em}
	\caption{The probability of error versus total number of released molecules, for ideal communication channel between FCs and gateway.}
	\label{fig: result 2}
	\vspace{0em}
\end{figure}

The probability of localization error versus the total number of molecules, $KN_{th}M$, is shown in Fig. \ref{fig: result 2}, in the presence of ideal communication channel between the FCs and the gateway. As can be seen, the non-collaborative sensors perform better then the collaborative ones when the number of released molecules is increased. Since the normal distribution used for the ratio of gateway observations is more accurate when the number of molecules is increased. Also we can see that the error probability increases for higher resolutions (higher values of $L$). Similar to Fig. \ref{fig: result 1}, the simulation results validates the numerical results.

\begin{figure}
	\vspace{-1em}
	\centering
	\includegraphics[trim={1cm 0cm 1cm 0cm},clip, scale=0.47]{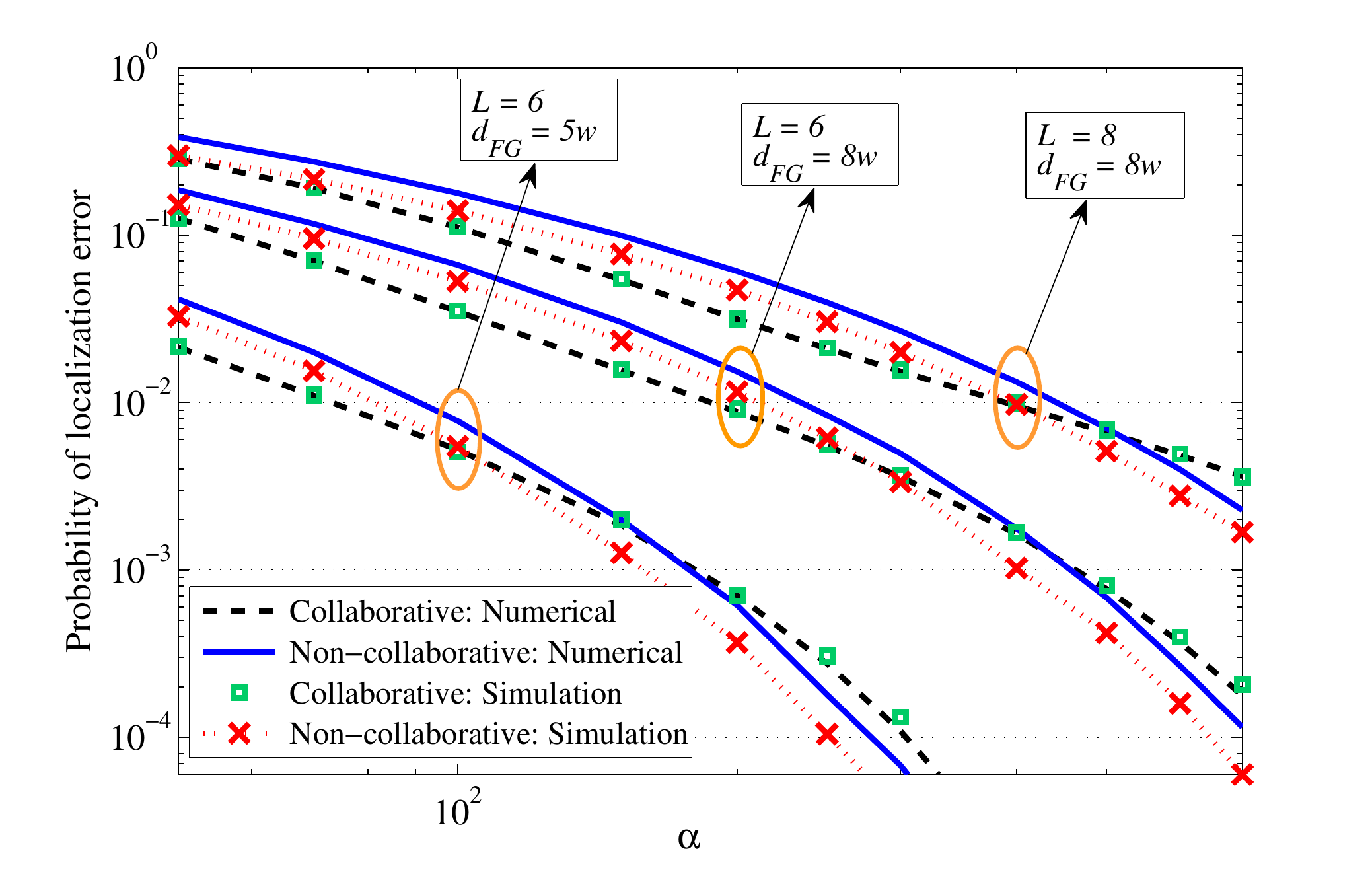}
	\vspace{-1em}
	\caption{The probability of error versus the amplification factor, for non-ideal channel between FCs and gateway, where $N_{th}M=10^8$}
	\label{fig: result 3}
\end{figure}

Now, we investigate the performance of the proposed model in the presence of non-ideal (noisy) communication channel between the FCs and gateway. 
The probability of localization error is plotted versus the amplification factor $\alpha$ for different distances between the FCs and gateway. As shown in Fig. \ref{fig: result 3}, the probability of localization error is decreasing versus the amplification factor. The performance of collaborative sensors is better for lower amplification factors. The error is higher for longer distances $d_{FG}$ and the simulation results validate the numerical results. 
{Also from Fig. \ref{fig: result 2} and Fig. \ref{fig: result 3}, it can be seen that the effect of $\alpha$ on the probability of error is similar to the effect of number of molecules ($N_{th}M$). Because, the decision rules and the probabilities of error depend on $m(d_i)$ for the case of ideal channel between the FCs and gateway, and $m_G(d_i)$ for the case of non-ideal communication channel. Reminding that $m(d_i) {=} N_{th} M \mu(d_i, \frac{w^2}{4ND})$ and $m_G(d_i)\simeq \alpha m(d_i)\tilde\mu(d_{FG}, T_{G,\text{obs}})$, the effect of $N_{th}M$ and $\alpha$ on $P_e$ would be similar.}

\begin{figure}
	\vspace{-1em}
	\centering
	\includegraphics[trim={1cm 0cm 1cm 0cm},clip, scale=0.47]{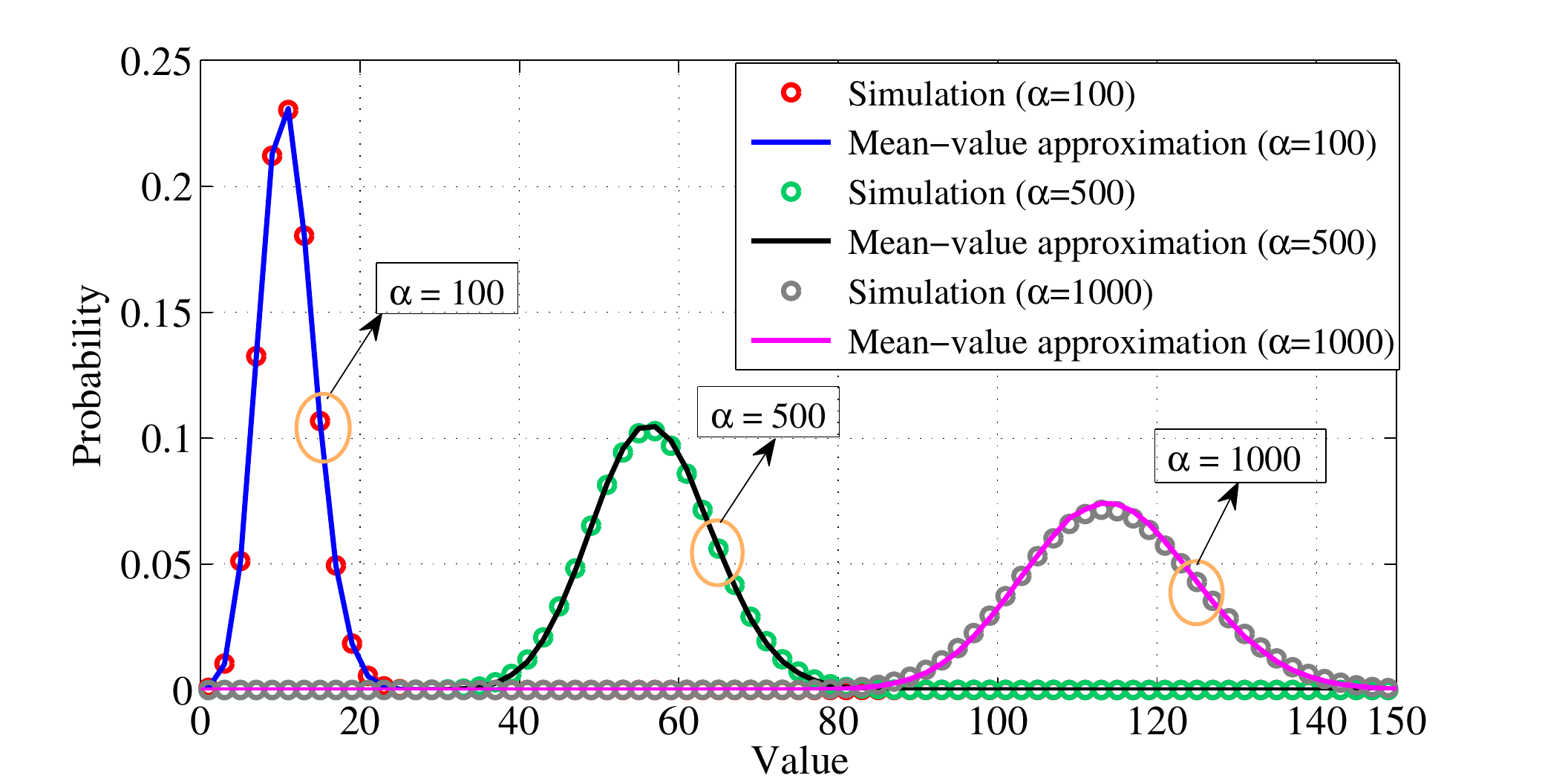}
	\vspace{-1.5em}
	\caption{The approximated PDF and the histogram of the observed signal at the gateway, where $d_1=8.3\times10^{-4}$ and $d_{FG}=5w$}
	\label{fig: result 4}
\end{figure}

In Fig. \ref{fig: result 4}, we plot the histogram and the approximated PDF of the number of markers observed at the gateway, for different values of amplification factor $\alpha$. The PDF is obtained by approximating the number of molecules observed at the FC with its mean value. As can be seen, two results are close to each other, which confirms the accuracy of the mean-value approximation used in Section \ref{sec: Non-ideal Communication Channel}. 

\section{Conclusion} \label{sec: conclusion}
	In this paper, we consider a general setup to perform joint
	sensing, communication and localization of a silent abnormality
	with molecular communication. We investigate the abnormality
	localization problem in a 2-D medium, with no physical access
	to the abnormality point, with three types of devices: the
	mobile sensors, FCs, and a gateway. We consider two types
	of collaborative and non-collaborative sensors, and study both
	cases of ideal and noisy communication channel between the
	FCs and gateway. For the collaborative sensors, we obtain
	the ML decision rule and sub-optimum thresholds, and derive
	the probability of error. For the non-collaborative sensors,
	we faced doubly stochastic RVs, which make our analyzes
	difficult. To overcome this problem, we use the ratio of gateway
	observations, and obtain the ML decision rule and probability
	of error in a closed form. For the case of noisy communication
	channel between the FCs and gateway, we use mean-value
	approximations for doubly stochastic RVs to obtain the decision
	rules and probabilities of error. It is observed that the noncollaborative sensors performs better that the collaborative ones
	when the number of molecules is increased. The proposed
	model can also be extended into a 3-D environment easily by
	utilizing one more FC.

\begin{appendices}
\section{Lemmas 1 and 2} \label{lemmas_1_and_2}
	\begin{lemma}{\cite[p. 262]{krishnamoorthy2006handbook}}:
		\label{lemma_1}
		If we have $K$ independent RVs $X_i \sim \mathcal{N}(\mu_i, \sigma^2_i)$, $i=1,\cdots,K$, then the RV $Z_i = \sum_{k=1}^{K}\left(\frac{X_i}{\sigma_i}\right)^2 $ has a non-central Chi-squared distribution with the following cumulative distribution function (CDF).
		$$P[Z_i<x]=1-Q_{\frac{K}{2}}(\sqrt{\lambda},\sqrt{x}),$$
		where $\lambda = \sum_{k=1}^{K}\left(\frac{\mu_i}{\sigma_i}\right)^2$, and $Q_M(a,b)$ is Marcum Q-function, defined as follows \cite{sun2008inequalities}.
		$${\displaystyle Q_{M}(a,b)=\int _{b}^{\infty }x\left({\frac {x}{a}}\right)^{M-1}\exp \left(-{\frac {x^{2}+a^{2}}{2}}\right)I_{M-1}(ax)\,dx},$$
		where $I_{M-1}$ is modified Bessel function of order $M-1$.
	\end{lemma}

	\begin{lemma}\cite{Gauss_Approx_2013}: \label{lemma_2}
	Let $V_1$ and $V_2$ be two independent RVs as $V_i \sim \mathcal{N}(\mu_i,{\sigma^2_i})$, $i=1,2$, such that $0<\sigma_1/\mu_1<\lambda \leq 1$ and $0<\sigma_2/\mu_2\leq \sqrt{\lambda^2-(\sigma_1/\mu_1)^2}<\lambda$, where $\lambda$ is a known constant. For any $Z$ that belongs to the interval $[\beta-\frac{\sigma_z}{\lambda}, \beta+\frac{\sigma_z}{\lambda}]$, where $\beta = \mu_1/\mu_2$, and $\sigma_z=\beta\sqrt{{(\frac{\sigma_1}{\mu_1})^2+(\frac{\sigma_2}{\mu_2})^2}}$, satisfies that $|G(z)-F_Z(z)|<\epsilon$, for every $\epsilon>0$, where $G(z)$ is the distribution function of a normal RV with mean $\beta$ and variance $\sigma_z^2$, and $F_Z(z)$ is the distribution function of $Z=V_1/V_2$.
\end{lemma} 
\end{appendices}

	\bibliographystyle{ieeetr}
	\bibliography{ref}

\end{document}